\documentclass{article}
\usepackage{amssymb}

\input{tcilatex}
\begin{document}

\author{ J. G. Cardoso \\
Department of Mathematics\\
Centre for Technological Sciences-UDESC\\
Joinville 89223-100, Santa Catarina, Brazil.\\
e-mail: dma2jgc@joinville.udesc.br
cosmic microwave background.}
\title{Wave Equations for Invariant Infeld-van der Waerden Wave Functions
for Photons and Their Physical Significance}
\date{ }
\maketitle

\begin{abstract}
The inner structure of the $\gamma \varepsilon $-formalisms of Infeld and
van der Waerden admits the occurrence of spin-tensor electromagnetic fields
which bear invariance under the action of the generalized Weyl gauge group.
A concise derivation of the wave equations for such fields is carried out
explicitly along with the construction of a set of torsionless
covariant-derivative expressions. It is emphatically pointed out that the
integration of the wave equations arising herein may under certain
circumstances produce significant insights into the situation concerning the
description of some physical properties of the cosmic microwave background.
\end{abstract}

\section{Introduction}

One of the striking features of the $\gamma \varepsilon $-formalisms of
Infeld and van der Waerden [1-3] is related to the fact that spacetime
curvature structures generally carry wave functions for photons of opposite
handednesses in an inextricable manner [3, 4]. Loosely speaking, such wave
functions amount in the case of either formalism to contracted curvature
contributions that enter locally into spinor decompositions of Maxwell
bivectors. Traditionally, the presence of intrinsically geometric
electromagnetic fields appears to be bound up with a single condition upon
the metric spinors for the $\gamma $-formalism, which remains formally
unaltered when the action of the generalized Weyl gauge group [5] is
effectively implemented. In spacetimes that admit nowhere-vanishing Weyl
spinor fields, background photons turn out to interact with underlying
gravitons. However, the relevant coupling patterns are strictly borne by the
wave equations that control the electromagnetic propagation [4].

Any electromagnetic curvature spinors for the $\gamma $-formalism behave as
spin tensors under the action of the Weyl group, whilst typical $\varepsilon 
$-formalism contributions are correspondingly looked upon as pairs of
gauge-invariant spin-tensor densities having appropriate weights and
antiweights. Remarkably enough, rearranging index configurations adequately,
produces the occurrence in both formalisms of wave functions for photons
that bear an invariant spin-tensor character. This notable property of the
formalisms was brought forward for the first time in Ref. [3]. It was
utilized thereabout mainly to facilitate setting out the entire system of
electromagnetic wave equations that should be tied in with the formalisms.

In the present paper, we derive in a concise way the wave equations for the
invariant spin-tensor fields we have just allowed for above, along with a
set of interesting covariant-derivative expressions. We believe that, under
the standard cosmological circumstances [6, 7], these wave equations may
enable one to gain fresh insights into the situation concerning the
description of some physical properties of the microwave background of the
universe. Actually, it is this aspect of our work which most strongly
ensures the relevance of the whole presentation.

We will have necessarily to recall in Section 2 the definitions and
prescriptions that are of immediate interest to us. The differential
expressions and wave equations are deduced together in Section 3. A few
remarks on the work are made in Section 4. Many of the conventions adopted
in Ref. [3] will be used throughout the work, but we shall occasionally
explain some of them. In particular, the effect on any index block of the
actions of the symmetry and antisymmetry operators is indicated by
surrounding the involved indices with round and square brackets,
respectively.

\section{Basic formulae}

Wave functions for geometric photons are defined in either formalism as%
\begin{equation}
\phi _{AB}\doteqdot \frac{i}{2}\omega _{ABC}{}^{C},\text{ }\phi _{A^{\prime
}B^{\prime }}\doteqdot \frac{i}{2}\omega _{A^{\prime }B^{\prime }C}{}^{C},
\label{e1}
\end{equation}%
with the $\omega $-spinors entering the contracted configurations%
\begin{equation}
\omega _{ABC}{}^{C}=\omega _{(AB)C}{}^{C}\doteqdot \frac{1}{2}W_{ABA^{\prime
}}{}^{A^{\prime }}{}_{C}{}^{C},  \label{e2}
\end{equation}%
and\footnote{%
In the $\varepsilon $-formalism, the $\phi $-fields taken up by Eqs. (1) are
gauge-invariant spin-tensor densities of weight and antiweight $-1$.}%
\begin{equation}
\omega _{A^{\prime }B^{\prime }C}{}^{C}=\omega _{(A^{\prime }B^{\prime
})C}{}^{C}\doteqdot \frac{1}{2}W_{A^{\prime }B^{\prime
}A}{}{}^{A}{}_{C}{}^{C}.  \label{e3}
\end{equation}%
The $W$-objects of Eqs. (2) and (3) may be thought of as arising from the
commutator structure%
\begin{equation}
\lbrack \nabla _{AA^{\prime }},\nabla _{BB^{\prime }}]\zeta ^{C}\doteqdot
W_{AA^{\prime }BB^{\prime }M}{}^{C}\zeta ^{M},  \label{e4}
\end{equation}%
with $\zeta ^{C}$ being an arbitrary spin vector, and $\nabla _{a}$ denoting
some torsion-free covariant-derivative operator for the formalism eventually
taken into consideration. In each formalism, the overall curvature spinors
of $\nabla _{a}$ are thus carried by the expansion\footnote{%
The parts $\omega _{AB(CD)}$ and $\omega _{A^{\prime }B^{\prime }(CD)}$
constitute the gravitational spin curvature for either $\nabla _{a}$ (for
further details, see Ref. [3]).}%
\begin{equation}
W_{AA^{\prime }BB^{\prime }CD}=M_{A^{\prime }B^{\prime }}\omega
_{ABCD}+M_{AB}\omega _{A^{\prime }B^{\prime }CD},
\end{equation}%
where the kernel letter $M$ stands for either $\gamma $ or $\varepsilon $.

The electromagnetic curvature spinors for both formalisms are taken to
fulfill the relationships%
\begin{equation}
\omega _{ABC}{}^{C}=2i\nabla _{(A}^{C^{\prime }}\Phi _{B)C^{\prime }},\text{ 
}\omega _{A^{\prime }B^{\prime }C}{}^{C}=2i\nabla _{(A^{\prime }}^{C}\Phi
_{B^{\prime })C},  \label{e5}
\end{equation}%
together with the prescriptions%
\begin{equation}
F_{AA^{\prime }BB^{\prime }}=-\hspace{0.02cm}(M_{AB}\nabla _{(A^{\prime
}}^{C}\Phi _{B^{\prime })C}+M_{A^{\prime }B^{\prime }}\nabla
_{(A}^{C^{\prime }}\Phi _{B)C^{\prime }}),  \label{e6}
\end{equation}%
and%
\begin{equation}
\frac{i}{2}W_{abC}{}^{C}=F_{ab}=2\nabla _{\lbrack a}\Phi _{b]}.  \label{e7}
\end{equation}%
It is shown in Ref. [3] that the formalisms usually bear the same quantity $%
\Phi _{a}$, which comes into play as an affine electromagnetic potential
that satisfies the Weyl principle of gauge covariance.

\section{Derivative expressions and wave equations}

The Weyl group has to be taken as one and the same in both formalisms.
Indeed, the only unprimed-index configuration that accounts for invariant
wave functions for photons in the $\gamma $-formalism is provided by $\phi
_{A}{}^{B}$. On the other hand, the spin-density character inherently
carried by the metric spinors for the $\varepsilon $-formalism [1], implies
that the respective field $\phi _{A}{}^{B}$ should be regarded as an
invariant spin-tensor wave function as well. Consequently, we can readily
write down in either formalism the simple expression%
\begin{equation}
\nabla _{a}\phi _{A}{}^{B}=\partial _{a}\phi _{A}{}^{B}-\vartheta
_{aA}{}^{C}\phi _{C}{}^{B}+\vartheta _{aC}{}^{B}\phi _{A}{}^{C},  \label{e8}
\end{equation}%
where $\vartheta _{aA}{}^{C}$ accordingly denotes an admissible spin
affinity. It follows that, manipulating the indices of the $\vartheta $%
-terms of Eq. (9), after some easy calculations, we obtain%
\begin{equation}
\nabla _{a}\phi _{A}{}^{B}=\partial _{a}\phi _{A}{}^{B}-\vartheta
_{a(AC)}{}M^{BD}\phi _{D}{}^{C}+\vartheta _{a}{}^{(BC)}\phi _{A}{}^{D}M_{DC},
\label{e9}
\end{equation}%
with the kernel letter $M$ bearing the same meaning as before. In fact, the $%
\varepsilon $-formalism versions of $\vartheta _{a(BC)}{}$ and $\vartheta
_{a}{}^{(BC)}$ show up, respectively, as invariant spin-tensor densities of
weights $-1$ and $+1$, whence both of the $\vartheta $-pieces of Eq. (10)
really bear gauge invariance in either formalism.

In both formalisms, the field equation for $\phi _{A}{}^{B}$ must be spelt
out as the massless-free-field statement%
\begin{equation}
\nabla ^{AB^{\prime }}\phi _{A}{}^{B}=0.  \label{e10}
\end{equation}%
Therefore, making use of the splitting [3]%
\begin{equation}
\nabla _{A^{\prime }}^{C}\nabla ^{AA^{\prime }}\phi _{A}{}^{B}=\Delta
^{AC}\phi _{A}{}^{B}-\frac{1}{2}M^{AC}\square \phi _{A}{}^{B},  \label{e11}
\end{equation}%
and taking account of the (symmetric) derivative\footnote{%
In both formalisms, the operator $\Delta ^{AB}$ amounts to $\nabla
_{C^{\prime }}^{(A}\nabla ^{B)C^{\prime }}$. It behaves in the $\varepsilon $%
-formalism as an invariant spin-tensor density of weight $+1$.}%
\begin{equation}
\Delta ^{AB}\phi _{A}{}^{C}=\frac{R}{6}M^{BD}\phi _{D}{}^{C}-\omega
^{(ABCD)}\phi _{A}{}^{H}M_{HD},  \label{e12}
\end{equation}%
with $R$ being the pertinent Ricci scalar, we arrive at the equation%
\begin{equation}
(\square +\frac{R}{3})\phi _{A}{}^{B}=(-2)\Psi _{AD}{}{}^{BC}\phi _{C}{}^{D},
\label{e13}
\end{equation}%
with the definitions%
\begin{equation}
\Psi _{ABCD}\doteqdot \omega _{(ABCD)},  \label{e14}
\end{equation}%
and%
\begin{equation}
\square \doteqdot \nabla _{CC^{\prime }}\nabla ^{CC^{\prime }}.  \label{e15}
\end{equation}%
The $\Psi $-spinor of Eq. (15) represents one of the wave functions for
gravitons [6-8]. It should be stressed that the legitimacy in either
formalism of Eq. (14), stems essentially from the common invariant
spin-tensor character of $\phi _{A}{}^{B}$.

Any procedures associated to the implementation of specific techniques for
solving Eq. (14) would become considerably simplified if the situations
being entertained were set upon conformally flat spacetimes. Under such a
circumstance, one would just deal in either formalism with%
\begin{equation}
(\square +\frac{R}{3})\phi _{A}{}^{B}=0.  \label{e16}
\end{equation}%
At this stage, we could call upon the invariant distributional methods of
Ref. [9] to treat systematically Eq. (17) in conjunction with the standard
Friedmann-Robertson-Walker cosmological model [7]. We will probably
elaborate upon this issue elsewhere.

\section{Conclusions and outlook}

Within manifestly cosmological frameworks, it might be expected on the basis
of the invariant geometric character of either $\phi _{A}{}^{B}$ that the
key properties of the cosmic microwave background should be naturally
described by solutions of Eq. (17) subject to suitably prescribed boundary
conditions. In this connection, the following expression for the
energy-momentum tensor for either formalism%
\[
T_{AA^{\prime }}{}^{BB^{\prime }}{}=\frac{1}{2\pi }\phi _{A}{}^{B}\phi
_{A^{\prime }}{}^{B^{\prime }},
\]%
would presumably be helpful for achieving present-time values of the energy
and linear momentum of the radiation.

We should observe that the right-hand side of Eq. (10) supplies elementary
index-displacement rules for covariant derivatives of gauge-invariant fields
for both formalisms. In practice, one could additionally invoke the useful
relation%
\[
\vartheta _{a(BC)}=\frac{1}{2}(S_{(B}^{bD^{\prime }}\partial _{C)D^{\prime
}}g_{ab}+S_{b(B}^{D^{\prime }}\partial _{\mid a\mid }S_{C)D^{\prime }}^{b}), 
\]%
which absorbs some $\gamma \varepsilon $-connecting objects along with a
spacetime metric tensor carrying the signature $(+---)$.

\end{document}